\documentclass[preprint]{aastex}
\usepackage{lineno}
\usepackage{epsfig}
\usepackage{graphicx}
\pagewiselinenumbers

\usepackage{natbib}

\newcommand{\egret}{\emph{EGRET}}
\newcommand{\fermi}{\emph{Fermi}}

\newcommand{\magic}{\emph{MAGIC}}
\newcommand{\veritas}{\emph{VERITAS}}
\newcommand{\Fermi}{\emph{Fermi}}

\newcommand{\Av}{A_\mathrm{V}}

\shorttitle{Fermi-LAT observation of SNR IC~443}
\shortauthors{Fermi-LAT Collaboration}

\begin{document}
\pagestyle{empty}
\title{Observation of Supernova Remnant IC~443 with \\ the Fermi Large Area Telescope}
\author{
A.~A.~Abdo\altaffilmark{1,2}, 
M.~Ackermann\altaffilmark{3}, 
M.~Ajello\altaffilmark{3}, 
L.~Baldini\altaffilmark{4}, 
J.~Ballet\altaffilmark{5}, 
G.~Barbiellini\altaffilmark{6,7}, 
D.~Bastieri\altaffilmark{8,9}, 
B.~M.~Baughman\altaffilmark{10}, 
K.~Bechtol\altaffilmark{3}, 
R.~Bellazzini\altaffilmark{4}
% ...
}
\altaffiltext{1}{Space Science Division, Naval Research Laboratory, Washington, DC 20375, USA}
\altaffiltext{2}{National Research Council Research Associate, National Academy of Sciences, Washington, DC 20001, USA}
\altaffiltext{3}{W. W. Hansen Experimental Physics Laboratory, Kavli Institute for Particle Astrophysics and Cosmology, Department of Physics and SLAC National Accelerator Laboratory, Stanford University, Stanford, CA 94305, USA}
\altaffiltext{4}{Istituto Nazionale di Fisica Nucleare, Sezione di Pisa, I-56127 Pisa, Italy}
\altaffiltext{5}{Laboratoire AIM, CEA-IRFU/CNRS/Universit\'e Paris Diderot, Service d'Astrophysique, CEA Saclay, 91191 Gif sur Yvette, France}
\altaffiltext{6}{Istituto Nazionale di Fisica Nucleare, Sezione di Trieste, I-34127 Trieste, Italy}
\altaffiltext{7}{Dipartimento di Fisica, Universit\`a di Trieste, I-34127 Trieste, Italy}
\altaffiltext{8}{Istituto Nazionale di Fisica Nucleare, Sezione di Padova, I-35131 Padova, Italy}
\altaffiltext{9}{Dipartimento di Fisica ``G. Galilei", Universit\`a di Padova, I-35131 Padova, Italy}
\altaffiltext{10}{Department of Physics, Center for Cosmology and Astro-Particle Physics, The Ohio State University, Columbus, OH 43210, USA}
% ...


\begin{abstract}
We report observation of the supernova remnant IC~443
(G189.1+3.0) with the {\it{Fermi Gamma-ray Space Telescope}} 
Large Area Telescope (LAT) in the energy band between 200~MeV and 50~GeV.
IC~443 is a shell-type supernova remnant with mixed morphology
located off the outer Galactic plane
where high-energy emission has been detected 
in the X-ray, GeV and TeV gamma-ray bands.
Past observations suggest IC~443 has been interacting
with surrounding interstellar matter. Proximity between dense shocked molecular 
clouds and GeV$-$TeV gamma-ray emission regions detected by \egret, \magic\ 
and \veritas\ suggests an interpretation 
that cosmic-ray (CR) particles are accelerated by the SNR.
With the high gamma-ray statistics and broad energy coverage
provided by the LAT, we accurately characterize the gamma-ray 
emission produced by the CRs accelerated at IC~443. 
The emission region is extended in the energy band with 
$\theta_{68} = 0.27^\circ \pm 0.01^\circ (stat) \pm 0.03^\circ (sys)$ 
for an assumed 2-dimensional Gaussian profile and overlaps almost completely with the 
extended source region of \veritas. Its centroid is displaced significantly from 
the known pulsar wind nebula (PWN) which suggests the PWN is not the major contributor 
in the present energy band. The observed spectrum changes its power-law slope continuously 
and continues smoothly to the \magic\ and \veritas\ data points. 
The combined gamma-ray spectrum (200~MeV $<E<$ 2~TeV) is reproduced well 
by decays of neutral pions produced by a broken power-law proton spectrum with a break 
around 70~GeV.  
\end{abstract}

\keywords
{gamma-rays: general, supernovae: individual (IC~443)}

%% for the journal to which you are submitting your paper o determine


\maketitle

\section{Introduction}
IC~443 is a well-studied supernova remnant (SNR), 
possessing strong molecular line emission regions that make it a case 
for a SNR interacting with molecular clouds. The SNR is one of the best candidates 
for revealing the connection among SNRs, molecular clouds and high-energy gamma-ray sources
as reviewed by \citet{Torres03}.

IC~443 is located in the outer Galactic plane 
and listed as a core-collapse supernova remnant (SNR), G189.1+3.0,
in Green's catalog \citep{Green04}.
The SNR has an angular extent of $\sim$45$^\prime$ in the radio
with a complex shape consisting of two half-shells with different radii (Shells A and B)
\citep[e.g.,][ and references therein]{Fesen80,BraunStrom86a, BraunStrom86b,Petre88, 
Furst90, Leahy04}.
Its age is uncertain: some analyses indicate a young age ($3-4$ ky)
\citep[e.g.,][]{Petre88,Troja08} but others indicate that it is older ($20 - 30$ ky)
\citep[e.g.,][]{Lozinskaya81, Chevalier99, Olbert01, Gaensler06, Bykov08, JJLee08}.
Its distance has not been measured directly but is assumed to be $\sim 1.5$~kpc, the distance
to the Gem OB1 association to which the SNR belongs
\citep[e.g.,][]{Woltjer72, Olbert01, Welsh03, Gaensler06}.
A pulsar wind nebula (PWN), CXOU~J061705.3+222127, has been found in the southern periphery of the SNR
but its association with the SNR has not yet been firmly established
\citep{Keohane97, Olbert01, Bocchino01, Leahy04, Gaensler06, Troja08}.
To this day pulsation has not been reported at the position of the putative pulsar. 

A general picture has been drawn from past observations and analyses
that a variety of dynamical processes are taking place in the complex structure of IC~443 
\citep[e.g.,][ and references therein]{Troja06,JJLee08,Troja08}.
The processes include: interaction of SNR shocks  
with molecular and atomic clouds of various densities which produced 
a break-out (Shell B) from Shell A as well as associated small-scale structures;
interaction of the half-shells with
another SNR G189.6+3.3 \citep[e.g.,][]{Asaoka94, Keohane97};
penetration of shock fronts into dense molecular clouds
leading to molecular line emission \citep[e.g.,][]{Denoyer79a,Denoyer79b,Denoyer81,
Huang86, Burton88, vanDishoeck93, Richter95, Chevalier99, Hewitt06};
and interaction between the PWN and the environment
\citep{Olbert01, Leahy04, Gaensler06, Troja08}.

Of special interest for this study are the detections of high and very high energy (VHE) gamma rays
in the IC~443 vicinity. EGRET detected a gamma-ray source above 100 MeV, 
co-spatial \footnote{We assume that the gamma-ray sources detected in the region are associated 
with locally accelerated CRs based on the spatial overlap with the IC~443 structure seen 
in the radio, IR, optical and X-ray bands.} 
with the SNR (3EG J0617+2238) \citep{Sturner95,Esposito96,Lamb97,Hartman99}. 
The MAGIC telescope discovered a VHE source, MAGIC~J0616+225 
\citep{Albert07} which is displaced with respect to the position of the EGRET source, 
and co-spatial with what appears to be the most massive molecular cloud in the neighborhood 
detected in $^{12}$CO and $^{13}$CO emission lines \citep{Burton88,Dickman92,Dame01,Seta98}. 
{\it{VERITAS}} has confirmed the VHE emission (VER~J0616.9+2230) and resolved the source  
to be extended \citep{Acciari09}. The centroids of these 3 gamma-ray sources are 
displaced from that of the PWN.

The LAT data for IC~443 provide an exciting opportunity to study 
the interaction of an SNR with the interstellar medium, 
cosmic-ray (CR) acceleration and subsequent
injection to the Galactic space.
The entire Milky Way has been deeply observed by the LAT and modeling 
of the diffuse emission thereon allows the emission associated with IC~443 
(the ``IC~443 contribution") to be considered 
separately from the underlying Galactic diffuse emission, 
which has contributions from inverse Compton scattering of CR electrons 
(the ``Galactic IC component") and CR electron and proton interactions 
with interstellar nuclei (the ``Galactic CR contribution"). In the LAT data 
the spatial extension of the contribution from IC~443 can be measured 
along with its broad-band spectrum

This paper is organized in the following sections:
A brief description of the observation, event reconstruction and
gamma-ray selection is given in section~\ref{observation}.
The analysis procedure is explained in section~\ref{analysis} including
the instrument response function (IRF) and  
separation of the Galactic CR contribution, Galactic IC contribution, 
extragalactic emission and instrumental background.
We present results on the spatial extension and spectrum of 
the IC~443 contribution in section~\ref{ic443contribution}.
Discussion is given in section~\ref{discussion} and the paper is concluded
in section~\ref{conclusions}.

\section{Observations, Event Reconstruction and Gamma-Ray Selection}
\label{observation}

\subsection{Observation in the Survey Mode}\label{surveymode}

The {\it{Fermi Gamma-ray Space Telescope}}, launched
on 11 June 2008, has been surveying all sky with the Large Area
Telescope (LAT) since August 2008.
Its wide field of view ($\sim 2.4$~sr), large effective area
($\sim 8000$~cm$^2$ at $>1$~GeV), the improved point spread function (PSF) 
\footnote{The PSF is significantly 
different for gamma rays detected in the front and back portions of the tracker 
described in this section: 
the first and second of the two values separated by $/$ are for those detected in the front 
and back, respectively. We use $\theta_{68}$ and $\theta_{95}$ with superscripts $psf$, 
$error$ and $ext$ to quantify
the PSF, source localization error and source extension, respectively. 
The integrated probabilities in the 2-dimensional angular radii of $\theta_{68}$ 
and $\theta_{95}$ are 68\% and 95\%, respectively. For a symmetric 2D Gaussian 
distribution $\theta_{95}$ is $1.62 \times \theta_{68}$.} 
($\theta_{68}^{psf} \sim 0.6^\circ/0.9^\circ$ at $E=1$~GeV) and the broad energy coverage
(20~MeV $-$ 300~GeV) bring much improved sensitivity
and gamma-ray statistics  over its predecessor \egret\
\citep{Atwood09}.

The LAT is a pair-conversion telescope where a
gamma-ray is converted to an $e^+ e^-$ pair. Their trajectories are
recorded in the tracker
and the subsequent shower development are sampled
both in the tracker and calorimeter.
The tracker is surrounded by a segmented anti-coincidence detector
which is used to reject events induced by charged CRs
\citep{Atwood09, Calib09}.
The LAT PSF ($\theta_{68}^{psf}$) is determined at lower energies by 
multiple scattering in the tracker. At higher 
energies the PSF approaches to the limit given by the granularity of the tracker channels: 
it is $0.16^\circ/0.26^\circ$ at 5~GeV and $0.11^\circ/0.15^\circ$ at 10~GeV \citep{Atwood09}.

The LAT was operated in the nominal all-sky survey
during the present observation.  In the observation the instrument axis was
tilted from the zenith toward
the orbit's north and south poles by 35~deg or 39~deg on alternate orbits to
make sky coverage uniform.
The trigger rate, mostly on cosmic rays,
was $\sim 2.2$~kHz in average and varied between the maximum of
$\sim 5.0$~kHz and the minimum of $\sim 1.6$~kHz dependently 
on the geomagnetic cut-off rigidity.
On-board filtering reduced the event rate to $\sim 450$~Hz
for the downlink.  Data taking is disabled
during passages through the South Atlantic Anomaly \citep{Calib09}.

\subsection{Gamma-ray Selection}\label{gammaselection}

Gamma-ray candidates are defined in 3 classes on the gamma-ray probability,
background expected in orbit, current knowledge of the astronomical
gamma-ray fluxes, and performance of the LAT.
The \textit{Diffuse} class has the tightest background rejection
of the three \citep{Atwood09}. However the effective area becomes small 
and strongly dependent energy below 200~MeV. 
The averaged rate for the {\it{Diffuse}} class event was $\sim 0.6$~Hz during
the observation.

In the survey mode the Earth limb, an extremely bright source of gamma-rays,
comes near the edge of the field of view.
We have removed these gamma-rays with the reconstructed
zenith angles greater than $105^\circ$.

Cosmic-ray induced background in the \textit{Diffuse} class becomes comparable in
intensity to gamma-rays from the IC~443 region at energies below
$100-200$~MeV and above $50-100$~GeV.
The background consists of residual cosmic rays misclassified as gamma rays
and cosmic rays that convert in the passive material just outside of the LAT
without leaving a signal in the anti-coincidence detector.
We limit the energy range of this analysis between 200~MeV and 50~GeV 
where the effective area and the instrumental background is best undersood
\footnote{We are currently developing an improved event classification procedure 
to retain higher effective area at lower energies and to reduce background 
contaminations in the entire energy range.}.

The data analyzed here were obtained between 4 August 2008 and 4 July 2009.
The gamma rays in the circular region-of-interest (ROI) of radius 15$^\circ$
centered at the best-fit centroid of the IC~443 contribution 
to be determined in section~\ref{soclike} ($\ell =189.05^\circ$, $b= 3.03^\circ$) 
are selected for later ananlyses. We refer to this set of events as the data set: 
the key selections described here are summarized in Table~\ref{dataselection}.

Events in the data set are binned in energy at 13 logarithmic steps 
of $0.184$ starting from 200~MeV.
The matching energy-dependent exposure is calculated based 
on the orbit location, pointing direction, orientation
and live-time accumulation of the LAT.
The intensity is then calculated by dividing maps of counts with
maps of exposure in each energy bin.

\section{Analysis Procedure}\label{analysis}

The present analysis focuses on determination of 
the centroid and extension of the IC~443 contribution after separation of the Galactic 
CR contribution in the region. The latter will reflect the detailed spatial structure  
of the molecular clouds through pionic and bremsstrahlung interactions and 
potentially can be mistaken as a part of the IC~443 contribution. 
The uniform all-sky coverage of the LAT observation allows us to isolate 
the IC~443 contribution cleanly  
from all Galactic-scale contributions and determine its extension accurately. 
The \textit{Sourcelike} analysis has been designated specifically for this kind of analyses. 
Spectral analysis has been done with the LAT Science Tool \textit{gtlike} 
\footnote{Available from http://fermi.gsfc.gov/ssc/data/access/lat/BackgroundModels.html, 
the Fermi Science Support Center url for the 
Science Tools.} and has been cross-checked by \textit{Sourcelike}.  
We give a brief description of the Instrument Response Function (IRF) 
and \textit{Sourcelike} before proceeding to the fitting.   

\subsection{Instrument Response Function}\label{irf}

The spatial extension and spectral features of the gamma-ray emission
are studied by comparing the observation with predictions of
source models. Predictions are made by convolving
the spatial distribution and spectrum of the source models
with the IRF and the exposure for the observation. 

The IRF describes the overall performance of the instrument, 
event reconstruction and gamma-ray selection. 
In the \fermi\ LAT it has been formulated, before the launch, 
using an instrument simulation program \citep{Atwood09}. 
The simulation program has been calibrated against
beam test results \citep{Atwood09, Calib09} and 
the predicted IRF has been validated on several bright point sources
in the early operation phase.  

The variation in the trigger rate 
results in variation in the fraction of the trigger-enabled
time (the ``live-time fraction'') between $\sim 94.3$~\% to $\sim 81.5$~\%.
Besides lowering the live-time fraction and the exposure,
CR hits overlaid on a genuine gamma-ray track can reduce reconstruction
efficiency and lead to incorrect event selection.
The overall inefficiency has been found to
scale linearly with the loss in the live-time fraction with a coefficient that depends on energy.
The IRF used in the analysis, IRF P6\_V3\_Diffuse, has been corrected  
for inefficiency by, for example, $+23$~\%, $+16$~\% and $+12$~\% 
at 200~MeV, 500~MeV and 1~GeV, respectively.   

\subsection{Extension Analysis with \textit{Sourcelike}}
\label{soclike}

\begin{figure}
\centering
\epsscale{1.1}
\includegraphics[width=12cm]{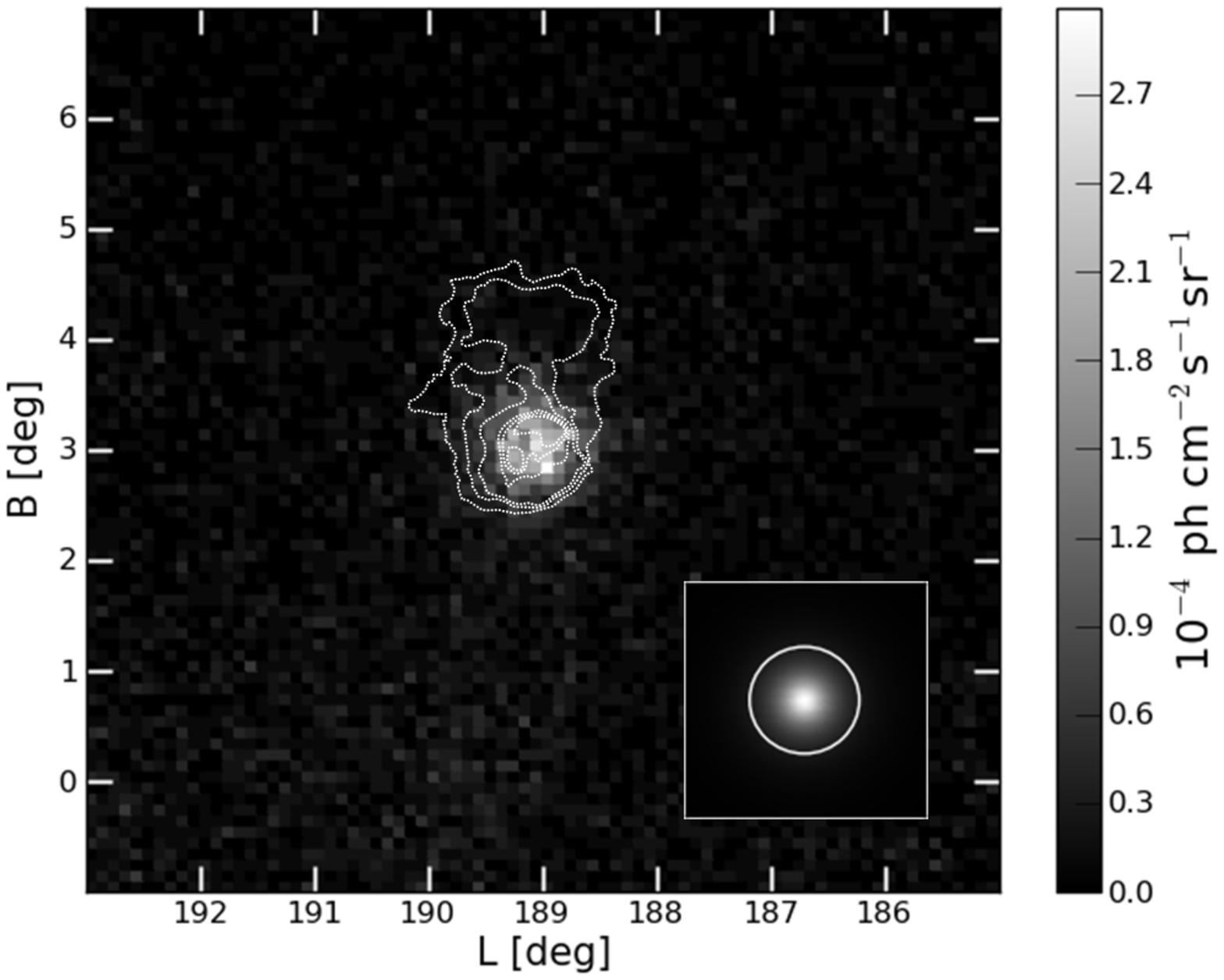}
\includegraphics[width=12cm]{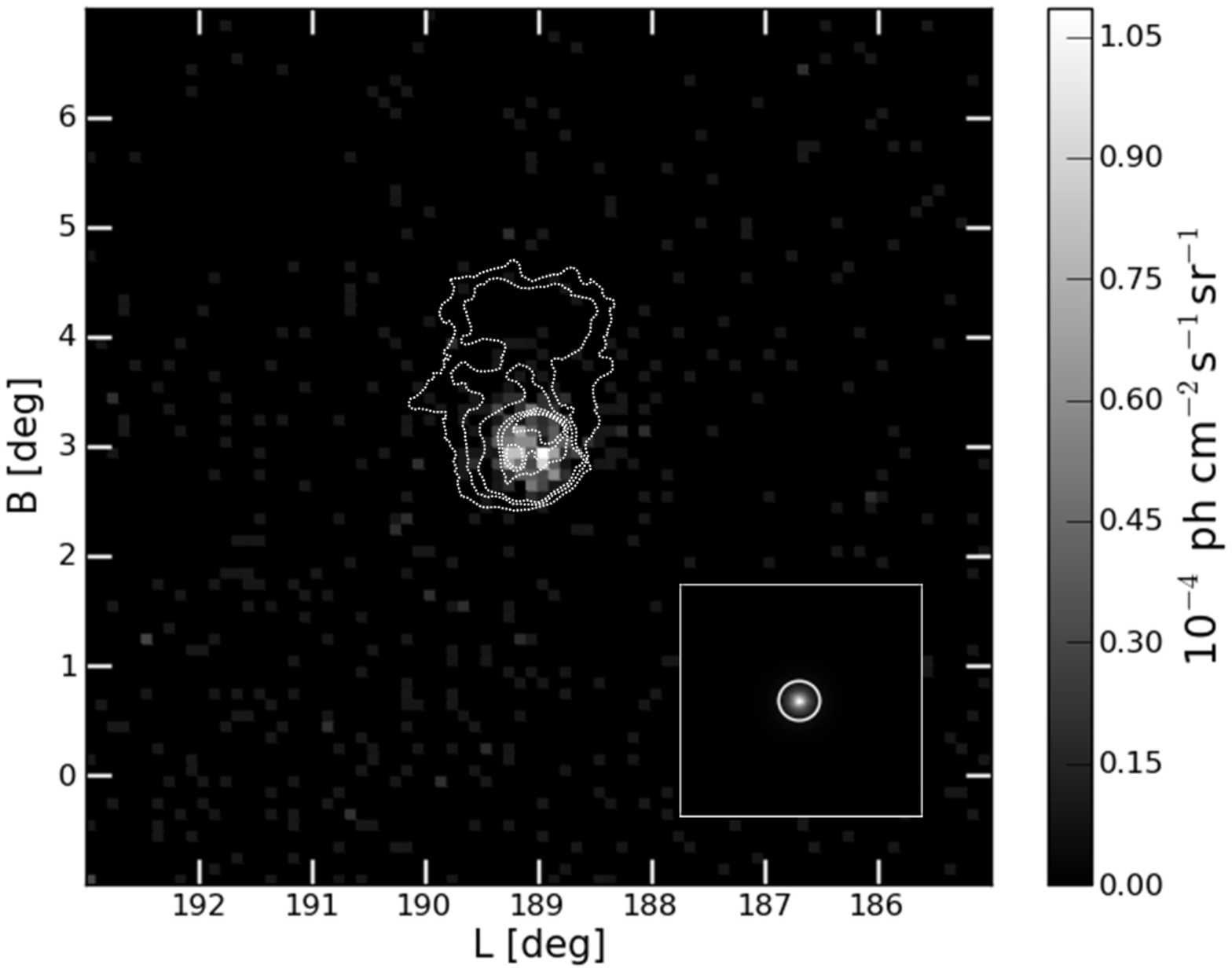}
\caption{Intensity map of the IC~443 region in the
$1-5$~GeV (left) and $5-50$~GeV (right) bands.
Units of intensity are $10^{-4}$~cm$^{-2}$~s$^{-1}$~sr$^{-1}$ for the color scale.
The overlay is the 2.7~GHz radio continuum contours taken from \citet{Furst90}.
The insets are the spectrum-weighted LAT PSF for each energy band, with the white circles
showing the corresponding $\theta_{68}^{psf}$. 
\label{intensitymaps}}
\end{figure}

The intensity distribution observed by LAT from the IC~443 region is 
shown in Fig.~\ref{intensitymaps} for a lower ($1 - 5$~GeV) and
a higher ($5 - 50$~GeV) energy bands for an area of $8^\circ \times 8^\circ$ centered
at ($\ell$, $b$) $=$ ($189^\circ, 3^\circ$) with $0.1^\circ$ pixelization. 
Spatial extension of the IC~443 contribution is determined on the intensity distribution 
using \textit{Sourcelike}, an analysis tool developed by the LAT team. 
In the tool, likelihood fitting is iterated to the data set 
assuming spatial source models and a spatial background model: 
we use combination of a symmetric 2D Gaussian source model or a point-source model 
and the standard background model.

The standard background model used in \textit{Sourcelike} is formulated by summing   
the Galactic CR contribution, Galactic IC contribution and isotropic 
component\footnote{The sum of the extragalactic background, unresolved sources
and instrumental background: its spatial distribution is assumed to be isotropic.} 
given in the diffuse emission model \footnote{\textit{gll\_iem\_v02.fit} 
and \textit{isotropic\_iem\_v02.txt} available from the url given in footnote 4.}: 
it is referred as the background here after. 
All bright sources detected with the LAT \citep{BrightSrc09} within 15$^\circ$ of the 
the centroid are included in the background. 

The fit is performed for the entire data in the user-determined energy range.   
Absolute normalization of individual background components can be constrained or 
unconstained in the fit: we leave the diffuse emission model as one unconstained 
component and so are all bright sources in the ROI in the \textit{Sourcelike} fit.  
The difference in Test Statistic (TS) values between the best-fit
Gaussian distribution and the best-fit point-source which is $2 \Delta log({\rm{Likelihood}})$
gives a measure of statistical significance of the extension. 
We refer to this difference as TS$_{ext}$ in this paper. 

The ROI is energy dependent in the \textit{Sourcelike}: $15^\circ$ at 200~MeV 
and shrinks to a minimum of $3.5^\circ$ at 50~GeV, which is at least a factor 
of $20$ larger than $\theta_{68}^{psf}$ of the LAT at the same energy and more than 
a factor of $10$ larger than the spatial extension ($\theta_{68}^{ext}$) of the source 
to be determined later.

\section{The IC~443 contribution}
\label{ic443contribution}

\subsection{Spatial Extension of the IC~443 Contribution}
\label{soclike}

Two energy bands, 1~GeV$<E<$5~GeV (the lower energy band) and 
5~GeV$<E<$50~GeV (the higher energy band) have been selected 
to study the spatial extension of the IC~443 contribution. 
\textit{Sourcelike} has been run for events in the two energy bands separately 
as well as in the combined energy band under a 2D Gaussian and point source hypotheses. 
The best-fit results are summarized in Table~\ref{extensiontable}. 

The number of gamma rays in the fitted 
Gaussian distribution is 4972 for 200~MeV$<E<$1~GeV, 1597 for 1~GeV$<E<$5~GeV  
and 236 for 5~GeV$<E<$50~GeV.  
For a given PSF, the accuracy of centroid determination is predicted to improve 
proportionally to the inverse of the square-root of the number of events. 
The accuracy quoted in Table~\ref{extensiontable} is consistent 
with this prediction for the effective PSF averaged over events in the 
energy bands 1~GeV$<E<$5~GeV and 5~GeV$<E<$50~GeV.
 
The difference in TS (TS$_{ext}$) between the symmetric 2D Gaussian and point hypotheses is 
$+106$ to $+121$ ($10.3$ to $11.0\ \sigma$)\footnote{The two TS$_{ext}$ values quoted are: 
the first one for that used in P6\_V3\_diffuse; and the second one for a worst-case PSF 
to be used later in section~\ref{discussion} to obtain a conservative systematic error.}  
for the $1-5$~GeV band and 
$+212$ to $+81$ ($14.6$ to $9.0 \sigma$)  
for the $5-50$~GeV band. The centroids for the two bands are consistent within $0.04^\circ$. 
The IC~443 contribution is extended to  
$\theta_{68}^{ext} = 0.26^\circ - 0.27^\circ$ in the two energy bands. 
The centroid in the high energy band is displaced southwards 
by $\sim 0.04^\circ$ ($\sim 1.5 \sigma$) in the Galactic coordinate from that 
in the low energy band.   

The results on source location and extension are robust: 
TS values have been examined at discrete points offset from 
the best-fit location and extension to confirm the fit.
To verify the fit further, we have generated 100 simulated sets of events 
assuming the best-fit centroid, extension and background 
with the LAT Science Tool \textit{gtobssim}. The simulated data are then  
processed through \textit{Sourcelike} under 2D Gaussian and point source 
hypotheses. The distribution of TS$_{ext}$ between the two hypotheses  
is consistent with the values given in Table~\ref{extensiontable}. 

The point source hypothesis is rejected at TS$_{ext}$$>81$ or $> 9\ \sigma$ 
independently in the two energy bands and at TS$_{ext}$$>212$ or $> 14\ \sigma$ 
in the combined energy band. 
The extensions in the two energy bands 
are mutually consistent within the errors given in Table~\ref{extensiontable}. 

\begin{figure}
\centering
\epsscale{1.1}
\plottwo{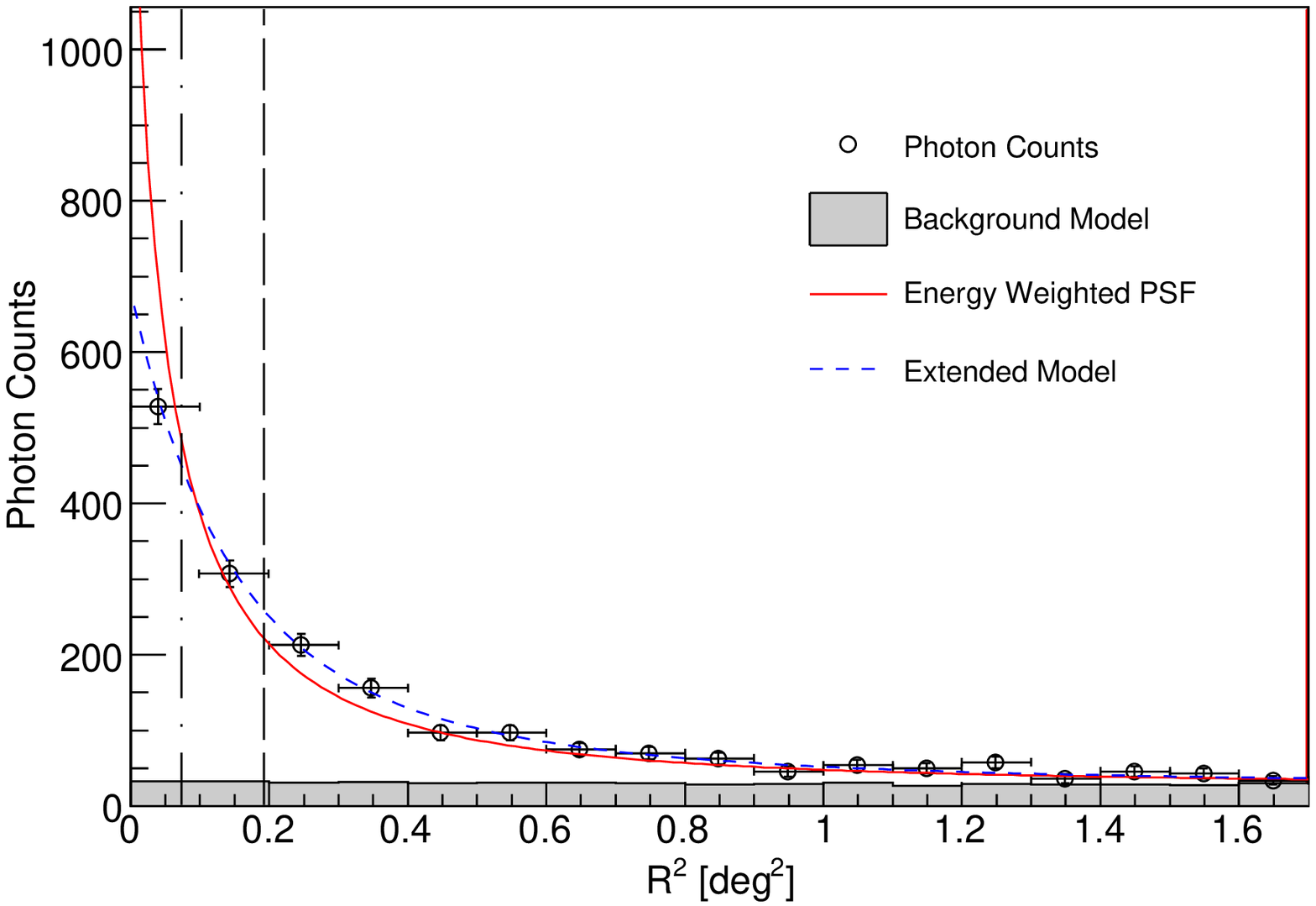}{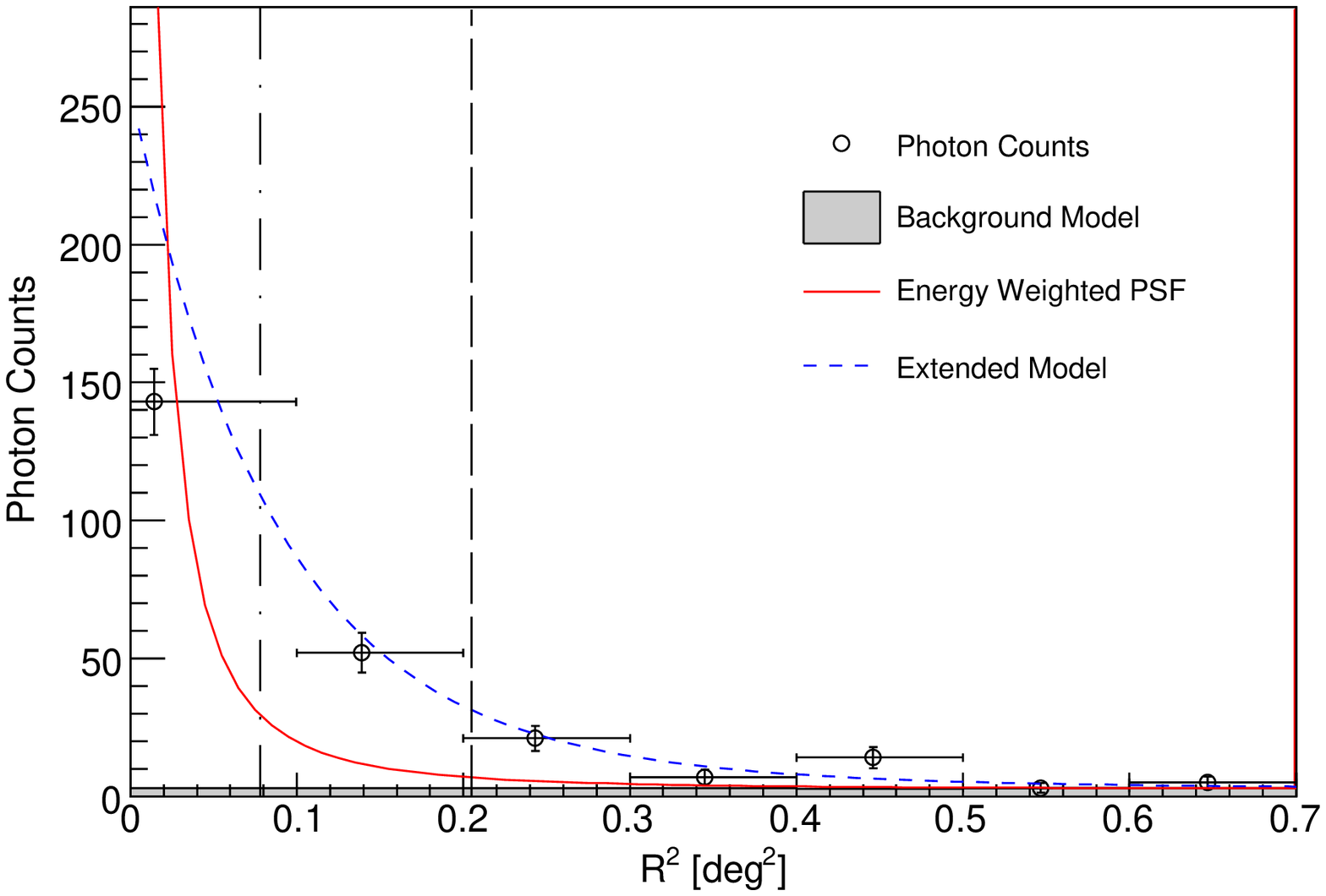}
\caption{Radial profile of the gamma rays \textit{SourceLike} associated 
with the IC~443 contribution (data points), compared with the effective spectrum-weighted LAT PSF 
(solid red line) and the fitted 2D Gaussian extended model (dashed blue line).
The left and right panels correspond to the low ($1-5$~GeV) 
and high ($5-50$~GeV) energy bands, respectively. The points in the count profile 
are plotted at the weighted average radial positions within their respective bins.
The vertical dash-dot and dashed lines correspond to the fitted extension 
$\theta_{68}^{ext}$ and $\theta_{95}^{ext}$ 
given in Table~\ref{extensiontable} for the respective energy bands.
\label{radialprofile}
}
\end{figure}

The radial profiles of event distribution around the centroid is shown 
in Fig.~\ref{radialprofile} for the low and high energy bands
together with the profile predicted for the point source hypothesis which is 
the LAT PSF weighted with the spectral distribution of the events 
\textit{Sourcelike} has associated with the source under the point-source hypothesis.

Extension was poorly determined for E=200~MeV $-$ 1~GeV because of the large PSF of the 
LAT in the energy range \citep{Atwood09}. However, the centroid and extension are consistent 
with the extension determined above 1~GeV and given in Table~\ref{extensiontable}. 
Hence we assume the same 2D Gaussian distibution in the entire energy range. 

\subsection{Spectrum of the IC~443 Contribution}
\label{fit}
The spectrum of the IC~443 contribution is fitted 
by the Science Tool \textit{gtlike}, the \fermi\ standard tool, as well as by  
\textit{Sourcelike}. In \textit{gtlike}, we have to assume a spatial template for all spectral 
components included in the fitting. The data 
set is assumed to be a sum of three contributions:
the best-fit 2D Gaussian distribution given for the $E=1-50$~GeV range  
in Table~\ref{extensiontable} which represents the IC~443 contribution \footnote{The extension 
could not be determined at a high statistical significance for $E=200$~MeV $-1$~GeV but 
the spatial distribution of gamma rays is consistent with those given 
in Table~\ref{extensiontable}.}; the 
background whose spatial distribution is represented by the sum of \textit{gll\_iem\_v02.fit} 
and \textit{isotropic\_iem\_v02.txt}; and the bright sources listed in \citet{BrightSrc09} 
in the square region of $8^\circ \times 8^\circ$ centered at the best-fit centroid 
($\ell = 189.05$, $b=3.03$).

The fitted IC~443 spectra from \textit{gtlike} and \textit{Sourcelike} 
agree well within the total error. We adopt the spectrum obtained with \textit{gtlike} 
and tabulate in Table~\ref{spectrumtable}. 
It is converted to the spectral energy density (SED) and  
shown by circles with error bars in Fig.~\ref{specfit}. 
Squares with error bars in the figure is the background spectrum 
normalized to the solid angle subtended by $\theta_{95}^{ext} = 0.45^\circ$ around the 
centroid given in Table~\ref{extensiontable}. 
The IC~443 contribution is approximately 20 times higher than the background in the entire 
energy band.

The SED of the IC~443 contribution thus determined has been fitted with a single power-law and 
broken-power-law models: the results are tabulated in Table~\ref{spectralfit}.
The single power-law fit fails to represent the spectrum 
giving a large reduced chi-square of $\sim 9$ while the broken power-law fit 
represents the overall shape quite well giving a small reduced chi-square 
($\sim 1.0$) as shown in Table~\ref{spectralfit}.  
The SED of the IC~443 contribution is plotted with those from previous observations,
\egret\ \citep{Esposito96}, \magic\ \citep{Albert07} and \veritas \citep{Acciari09} 
in Fig.~\ref{broadbandspectrum}.  

\subsection{Systematic error in determination of the centroid, extension and spectrum}\label{systematicerror}
 
When we determine the centroid of the IC443 contribution,
uncertainty in the spatial distribution of the Galactic diffuse emission adds
to the systematic error. The spatial template is taken from the standard diffuse emission 
model, \textit{gll\_iem\_v02.fit}. To confirm our analysis,
we have fitted the data set with the standard version of GALPROP for \fermi\ LAT 
(GALDEF 54\_59Xvarh8S) \citep{Strong98, Strong00, Strong09} made of the CO line survey 
by \citet{Dame01} and H~I survey by \citet{Kalberla05} as well as with a gas model made 
of the $\Av$ map by \citet{Dobashi05} and H~I survey by \citet{Kalberla05}.
The two alternate gas models have given centroids consistent with that given 
in Table~\ref{extensiontable}. 

The residual misalignment of the LAT and the star tracker can
also contribute to the systematic error: the source localization has been verified 
on orbit using bright point sources to $\pm 30$~arc-sec as of August 2009.
The combination of all errors described here gives the overall systematic 
localization error in Table~\ref{locationsummary}.

Our flux measurement depends on the knowledge on the effective area as a function of
gamma-ray energy. We estimate systematic error in the effective area 
to be 10~\%, 5~\%, and 20~\% at E=100~MeV, 562~MeV and $>10$~GeV respectively.

Uncertainty in the background used in \textit{gtlike} and \textit{Sourcelike} can also introduce 
error in the flux measurement. This uncertainty is  
estimated to be $\sim 20$\% for 200~MeV $-$ 1~GeV and  
$\sim 30$\% for $>$ 1~GeV of the background \citep{NoGeVExcess09, LocalHI09}. 
Systematic error at each energy bin is determined through a linear interpolation in $log_{10}(E)$   
among the values quoted above. 

The PSF used in this analysis has been derived on the detector simulation 
which was itself verified in accelerator tests
\citep{Atwood09}. As gamma-ray statistics improves, the PSF will be 
updated against measurement on bright point sources. In the present study, we have used a preliminary 
upper limit to assess possible systematic error 
introduced by inaccurate formulation of PSF: we consider this as the ``worst-case" PSF. 
The worst-case PSF ($\theta_{68}^{psf}$) gives a widest limit while the 
for $E > 5$ GeV is about 40\% larger than the default PSF. 
The source centroid comes out to be consistent within the total error 
when \textit{Sourcelike} is run with the worst-case PSF. 
We have included the difference in the systematic error given 
in Tables~\ref{extensiontable} and \ref{locationsummary}.           

\section{Discussion}
\label{discussion}

The IC~443 system consists of a complex distribution of molecular and atomic clouds
in the southern rim of Shell A \citep[e.g.,][ and references therein]{Snell05,JJLee08}.
Molecular clouds wrap around the southern rim and 
the boundary region between Shells A and B \citep{IRAS88,Seta98,Dickman92,Dame01}.
Molecular lines from shocked gas have been found 
in several clouds suggesting interaction with the blast-wave
at multiple sites \citep[e.g.,][]{Cornett77, Denoyer81, Huang86, Burton88, Dickman92,
vanDishoeck93, Richter95, Seta98, Snell05}.
A prominent band of HI gas has also been found in the southeastern
part of Shell A \citep[][ and references therein.]{Denoyer78,Giovanelli79,JJLee08}.
Some parts of the H~I gas are found to be shocked \citep{BraunStrom86a,JJLee08}.

The total mass of the molecular gas in the region is estimated to be $\sim 1\times 10^4$~M$_\odot$
\citep{Torres03}, of which only a small fraction is shocked \citep{Snell05}.
The total mass in the H~I belt is estimated at $\sim 730$~M$_\odot$ 
of which $\sim 500$~M$_\odot$ is shocked \citep{JJLee08}. 
Despite past extensive observations and
analyses, little is known about how the multiple shell-like 
structures are spatially correlated and where one or more supernova explosions took place.

In the group of shocked molecular clouds schematically shown in Fig.~\ref{relativelocation}, 
Cloud G \footnote{Labeling is given by
\citet{Denoyer79b,Huang86}.} lies 
closest to the centroid of the \magic\ and \veritas\ sources \citep{Huang86}.
It appears to be extended by $\sim 8^\prime$  
and overlaps with a non-shocked CO cloud formation \citep{Huang86,Burton88, Dickman92}.
\citet{Chevalier99} has suggested that interaction between Shell A and
Cloud G is responsible for the gamma-ray emission observed by \egret.
Existence of an OH Maser in the cloud suggests that the densities reach 
$\sim 10^4$~cm$^{-3}$ \citep{Frail96,Hewitt06,Hewitt08}.
A later CO line observation by \citet{Snell05} found a compact core 
of extension $\sim 1^\prime-2^\prime$ in Cloud G at ($\ell=189.03^\circ$, $b=2.90^\circ$).
Fainter maser emission has also been found in Clouds B and D at 
($\ell=189.18^\circ$, $b=2.97^\circ$) and ($\ell=189.25^\circ$, $b=3.13^\circ$), 
respectively \citep{Hewitt06, Hewitt08}.

The locations and extensions of the gamma-ray emission from IC~443 detected 
by \egret\ \citep{Hartman99}, \magic\ \citep{Albert07}, \veritas\ \citep{Acciari09}, 
and \Fermi\ LAT are summarized in
Table~\ref{locationsummary} and shown in Fig.~\ref{relativelocation}. 
\fermi\ LAT gives the best source localization or the smallest error circle ($\theta_{68}^{error}$) 
for the 2D Gaussian centroid ($0.03^\circ$) and a 
precise determination of the source extension 
($0.27^\circ \pm 0.01^\circ (stat) \pm 0.03^\circ$ (sys)). 
Our centroid determined for 1~GeV~$<E<$~50~GeV, ($\ell=189.05^\circ$, $b=3.03^\circ$), 
is 0.05~deg away from the \egret\ source (3EG~J0617+2238) 
but within $\theta_{95}^{error}$ of their localization error; 
0.15~deg from the \magic\ source (J0610+225) which is at more than 
5 times their localization error ($\theta_{68}^{error}$); and
0.12~deg from the \veritas\ source (VER~J0616.9+2230) or 
at 1.5 times their localization error ($\theta_{68}^{error}$).

The measured source extension, $\theta_{68}^{ext} = 0.27^\circ \pm 0.01^\circ (stat) 
\pm 0.03^\circ (sys)$, is comparable with  
$\theta_{68}^{ext} = 0.24^\circ \pm 0.05^\circ (stat) \pm 0.06^\circ (sys)$\footnote{We assume 
the extension is modeled by a symmetric 2D Gaussian and converted to 
$\theta_{68}=1.51 \times \theta_{1 \sigma}$.} given by \veritas. 
The two extended regions overlap almost completely.   
The three shocked clouds with OH maser (Clouds B, D and G) are within our 
measured extension and so are other shocked clouds (Clouds C, E, F and H).  
The PWN localized at ($\ell=189.227^\circ$, $b=2.897^\circ$) by \citet{Olbert01} and \citet{Gaensler06} 
is 0.26~deg away from our centroid but within our measured extension $\theta_{68}^{ext}$. 

\begin{figure}
\centering
\epsscale{0.6}
\plotone{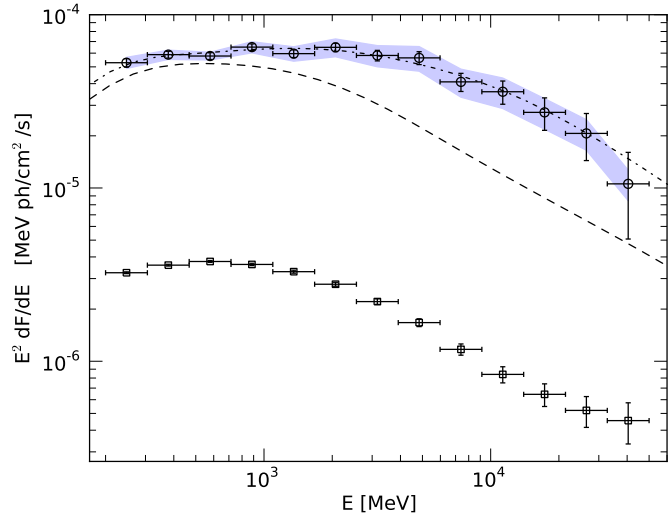}
\caption{The gamma-ray spectrum of the IC~443 contribution:
The upper and lower data points represent the IC~443 contribution 
and the total background, respectively.
The background has been scaled to match the solid angle subtended 
by a disk of radius $\theta_{95}^{ext} = 0.45$~deg. 
Errors are shown by the bars (statistical) and the grey band (systematic).  
The lines represent the pionic gamma-ray spectra produced 
by the Galactic CR proton at IC~443 scaled up by a factor of 100 (dashed) and 
by the locally accelerated proton population with the best-fit broken power-law spectrum 
described in the text (dot-dash). 
\label{specfit}
}
\end{figure}

The \fermi\ spectrum of the IC~443 contribution shown in Fig.~\ref{specfit} 
is flat between a few 100~MeV and $\sim 3$~GeV, suggesting the origin being mostly 
neutral pions produced by protons \footnote{We include cosmic-rays and target nuclei 
heavier than the proton (the alpha particle and heavier nuclei) in "protons" 
throughout this paper. In the approximation we adopt here \citep{Gaisser92}, 
these can be accounted for by multiplying a "nuclear factor" ($\sim 1.7$) 
without changing the CR proton spectrum. The known gamma-ray producing particle 
processes which do not go through neutral pions (e.g., $\eta^0 \rightarrow \gamma \gamma$ 
and direct photon production) contribute less than 1~\% in the present energy range.}. 
The dashed line in the figure represents the gamma-ray spectrum expected from
a $10^4$~M$_\odot$ cloud bombarded with the Galactic CR protons predicted at IC~443 
scaled up by a factor of 100. The Galactic CR spectrum is taken from the standard 
GALPROP (54\_59Xvarh8S) \citep{Strong98, Strong00, Strong09} and the 
parameterized cross-section for $pp \rightarrow \gamma$ by \citet{Kamae06}. 
In GALPROP the Galactic CR proton spectrum 
depends on the radius from the Galactic Center and the displacement from the 
Galactic Plane. The spectrum at the radius of IC~443 is $\sim 10$~\% lower 
than that in the solar vicinity and has a power-law shape with index $\sim 2.7$. 

\begin{figure}
\centering
\epsscale{1.4}
\includegraphics[width=12cm]{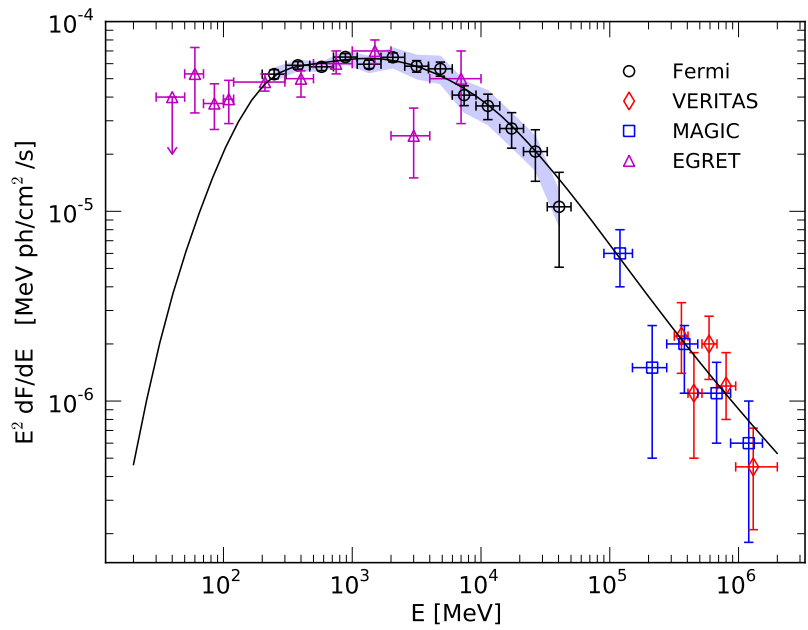}
\caption{Broadband spectral energy density of the 4 sources:  
\egret\ (purple triangles), \magic\ (blue squares), \veritas\ (red diamonds) and  
\fermi\ (black circles). The solid line is the same as the dot-dash line in Fig.\ref{specfit}.
The systematic and statistical errors of the Fermi data points 
are also the same as in Fig.~\ref{specfit}.
\label{broadbandspectrum}
}
\end{figure}

\begin{figure}
\centering
\epsscale{1.2}
\includegraphics[width=12cm]{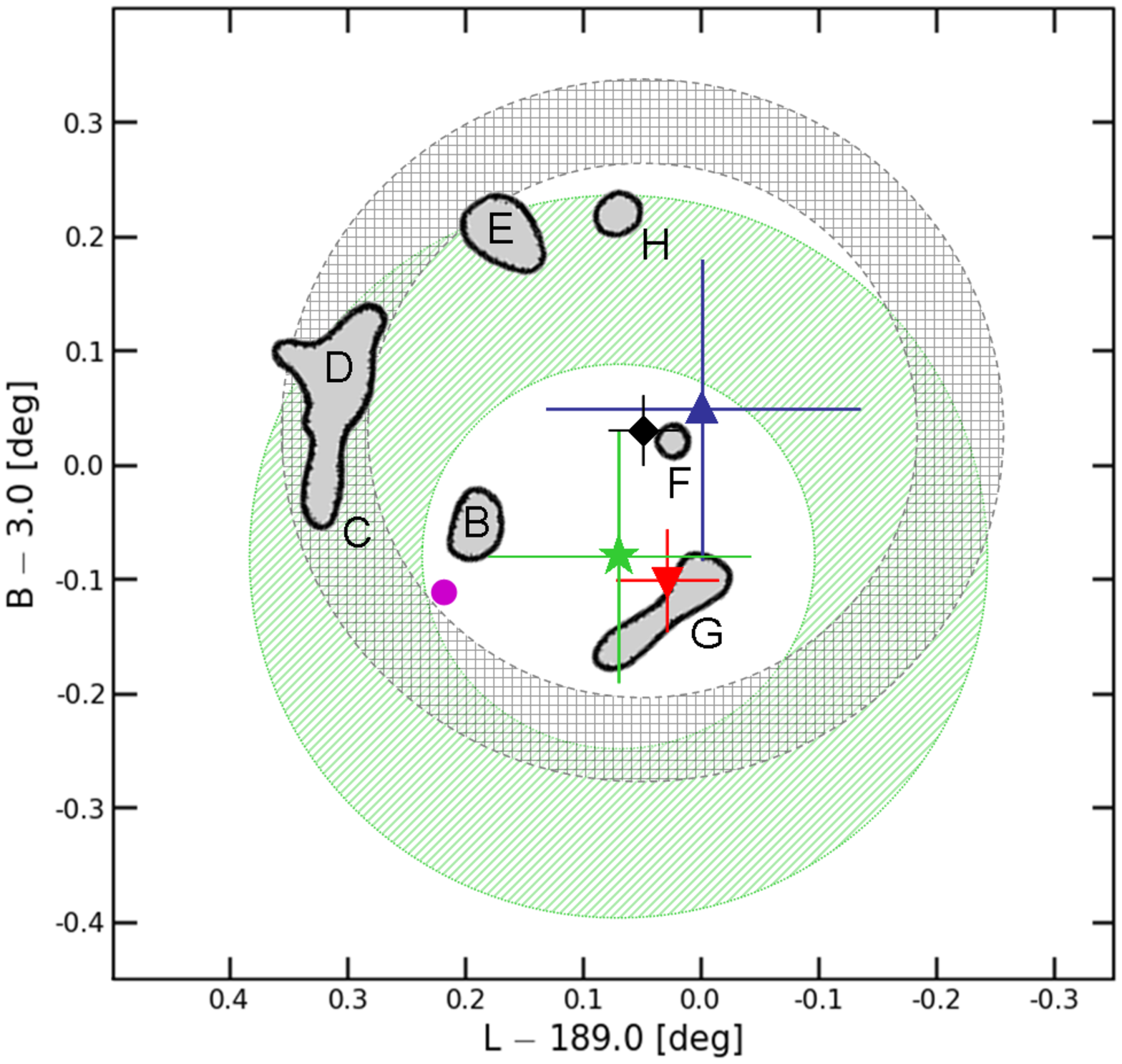}
\caption{Locations and extensions of the 4 gamma-ray sources: 
\egret\ centroid ($\triangle$); \magic\ centroid ($\bigtriangledown$); 
\veritas\ centroid (star) and \fermi\ LAT centroid ($\diamond$).
The respective localization errors as tabulated in Table~\ref{locationsummary} 
are shown as crosses. Best-fit spatial extensions of the \fermi\ (cross-hatched band) 
and \veritas\ (striped green band) sources are
drawn as rings with radii of $\theta_{68}^{ext}$ and widths of $\pm 1 \sigma$ error.    
The PWN location is shown as a dot. Contours are the locations and shapes 
of the local shocked molecular clouds taken from \citet{Huang86}.
\label{relativelocation}
}
\end{figure}

The \fermi\ SED is compared with those of \egret\, \magic\ and \veritas\ 
in Fig.~\ref{broadbandspectrum}.
The \egret\ spectrum is consistent with our spectrum except for their 3~GeV point.
\magic\ and \veritas\ do not overlap with the LAT in the energy coverage. Their fluxes are
consistent with ours if extrapolated down to $\sim 50$~GeV by assuming their measured power-law indices.

Since the source regions of \fermi\ LAT and \veritas\ overlap within their respective uncertainties 
listed in Table~\ref{locationsummary}, 
we can judiciously proceed to fit the 2 spectra with one spectral model. 

On the assumption that the distance is $d= 1.5$~kpc, the isotropic luminosity of IC~443 integrated over 
the energy band ($0.2 - 50$~GeV) is $1.2 \times 10^{35}$~erg/s. 
Electron bremsstrahlung can hardly explain the observed IC~443 gamma-ray emissivity: 
the cross-sections for bremsstrahlung and pionic gamma-ray emission are similar 
in the present energy band, so the bremsstrahlung-to-pion ratio is approximately 
the ratio of CR electron and proton fluxes which is $\sim 0.01$. 
The observed gamma-ray flux is too high for bremsstrahlung to be the dominant process. 
Inverse Compton scattering can not explain the observed IC~443 gamma-ray emission either: 
the gas density of the emission region is $\sim 50-100$~cm$^{-3}$ and 
the Compton-to-bremsstrahlung ratio is $\sim 0.01-0.001$ for the seed photon density 
of the cosmic microwave background. We note that there is no bright source of seed photons 
known in the region of the IC~443 contribution. The gamma-ray energy 
will be strongly bound by the electron spectrum which likely rolls down similarly  
as the proton spectrum. Our observation, however, does not rule out a small 
contribution from bremsstrahlung near the minimum of the present energy band.  

In a hadronic scenario, the observed photon spectrum up to TeV energies can be
well fitted by an underlying pion-producing proton population with 
a broken power-law spectrum F$_p$(T$_p$) = $5.9 \times 10^{-2}$~(T$_p/69$~GeV)$^{-\alpha}$
($10^4$M$_\odot /$M$_{gas}$)~cm$^{-2}$s$^{-1}$GeV$^{-1}$, 
where $\alpha$ is $2.09 \pm 0.04$ for T$_p < 69$~GeV and $2.87 \pm 0.07$ for T$_p > 69$~GeV respectively, 
and M$_{gas}$ is the gas mass in the interaction region. The error (statistical) in the fitted 
break energy is $\pm 25$~GeV and the chi-square for the best-fit broken power-law model
is $9.6/14$ per degree-of-freedom. Assuming the gas density (n~cm$^{-3}$) is uniform and 
the proton spectrum is the broken power-law everywhere in the interaction region, 
the total energy of the interacting protons is given by 
W$_p$($>0.5$~GeV) = $5.6 \times 10^{48}$(n$/240$~cm$^{-3}$)$^{-1}$(d$/1.5$~kpc)$^2$~erg.
Note that the pion production threshold is $\sim 0.5$~GeV. 
Taking M$_{gas} \sim 10^4$~M$_\odot$ and a gas volume ranging from 
$4\pi/3 \times$($\theta_{68}^{ext}$d)$^3$ = $5 \times 10^{58}$~cm$^3$ up to 
$4\pi/3 \times$($\theta_{95}^{ext}$d)$^3$ = $2 \times 10^{59}$~cm$^3$, 
we obtain n $\sim 60 - 240$~cm$^{-3}$, giving W$_p$($>0.5$~GeV)~$=(0.56-2.2) \times 10^{49}$~erg. 
We note that energies carried by local nuclear cosmic rays outside of the interaction region 
and by local leptonic cosmic rays are not included in the estimation. 
The fitted gamma-ray spectrum is shown in Fig.~\ref{specfit} 
and Fig.~\ref{broadbandspectrum}. We note that inclusion of the \magic\ points in the fit 
does not change the above results.

Broad-band gamma-ray spectral models have been proposed assuming CR interaction with
interstellar gas in IC~443 by \citet{Torres08} and \citet{Zhang08}. 
\citet{Torres08} model the CR diffusion in the SNR to allow spectral differences 
in the protons interacting with the ambient gas in the shell and in a detached molecular 
cloud. \citet{Zhang08} predict one contribution to come from 
the SNR shell evolving in the interstellar matter and 
the other from CR interaction with molecular clouds. One difference between the 
two models is that \citet{Zhang08} include inverse Compton scattering as a possible emission mechanism. 

The combined spectrum of \fermi\ and \veritas\ gives a strong constraint to 
spectral models for the IC~443 contribution. Since the spectrum of the dominant CR component 
(proton) is rolling over at $\sim 70$~GeV, 
secondary electrons and positrons can only contribute at energies   
$E_\gamma < 7$~GeV. This constrains the parameter space of the model by \citet{Zhang08}. 
The overlap between the \fermi\ and \veritas\ spatial extensions and the smooth spectral 
transition from \fermi\ to \veritas\ constrain the parameter spaces of 
the models by \citet{Torres08} and \citet{Zhang08}.      

We discuss briefly about possible mechanisms behind the broken power-law form of the proton
spectrum deduced from the gamma-ray observation of IC~443. 
The most obvious one is escape of highest energy CRs 
from the acceleration site. When accelerated CR protons exceed the maximum energy 
determined by the magnetic field and linear-scale of the aceleration site, 
they escape into the Galactic space. Theory of diffusive shock 
acceleration (DSA) assumes spherically symmetric morphology and predicts 
the CR spectrum to roll over exponentially at the maxmum energy. 
The maximum energy depends on the condition 
of the acceleration site: \citet{Ptuskin05} have incorporated various instabilities in DSA 
and predict the maximum energy as a function of the SNR age. 
For the age of IC~443 ($\sim 30$~kyr), the maximum proton energy can be around 
100~GeV \citep[Fig.1 of ][]{Ptuskin05} or near the observed break energy ($\sim 69$~GeV)  
beyond which the spectrum is assumed to cut-off exponentially. 
We have fitted the observed gamma-ray spectrum with 
a single power-law, exponentially cut off proton spectrum  
to get chi-square per degree-of-freedom 
of 30.3/15 much higher than 9.6/14 for the broken power-law spectrum. 
This simple statistical test therefore suggests that the observed broad-band gamma-ray 
spectrum is inconsistent with the simple DSA-based CR escape scenario which predicts a 
simple powerlaw with an exponential cutoff in the proton spectrum. We also note that 
DSA has mostly been studied for uniform gas densities around 1cm-3, while the gas 
around IC443 is inhomogenous and in various shocked molecular clouds around the 
remnant also denser.

Historically, after the discoveries of the \egret\ source \citep{Sturner95,Esposito96} 
and the hard X-ray source \citep{Keohane97} but before the discoveries     
of the PWN \citep{Olbert01} and \magic\ source \citep{Albert07}, models have been proposed 
to explain the emission between $\sim 5$~keV and $\sim 5$~GeV by bremsstrahlung in dense clouds 
\citep[e.g., ][]{Bykov00} with possible mix of synchrotron \citep[e.g., ][]{Sturner97}.
The SNR was also studied as a part of non-linear shock evolution in  
various environments \citep[e.g., ][]{Baring99}. While their predictions for IC~443 are not 
supported by the later observations including the present one, parameters in 
these studies can be readjusted to describe the bremsstrahlung contribution discussed below. 
 
The bremsstrahlung likely makes a non-negligible contribution below $E_\gamma =200$~MeV where 
the \egret\ data points exceed the best-fit pionic spectrum (see Fig.~\ref{broadbandspectrum}).
As our understanding of the IRF and cosmic-ray-induced background improve, analysis will be
extended to energies lower than 200~MeV and the bremsstrahlung spectrum component will 
be determined accurately. The hard X-ray SED measured by Beppo-SAX \citep{Bocchino00} 
is substantially higher than that by XMM \citep{Bocchino01}: 
which may suggest bremsstrahlung contribution 
near the PWN location as has been discussed by \citet{Sturner97} and \citet{Bykov00}.   

\section{Conclusions}\label{conclusions}

We have studied gamma-ray emission from the nearby SNR IC~443 (G189.1+3.0)
using the first 11 months of science data from \fermi\ LAT. The uniform sky coverage and 
high gamma-ray statistics of the observation have enabled us to separate 
the genuine IC~443 contribution from       
the emissions by Galactic CRs on interstellar gas, inverse Compton scattering 
by Galactic CR electrons on large-scale interstellar radiation field, 
extragalactic sources and instrumental background. 

Based on the extension study described in subsections~\ref{soclike} and 
the spectral analysis described in subsection~\ref{fit}  
as well as discussions given in section~\ref{discussion}, we conclude that:
\begin{itemize} 
\item The gamma-ray emission from IC~443 is detected at $\sim 86 \sigma$ level: 
the emission is extended with 68\% containment angular radius 
$\theta_{68}^{ext} = 0.27^\circ \pm 0.01^\circ \pm 0.03^\circ$ 
in the energy range between 1~GeV and 50~GeV. The extension remains unchanged  
within error in the low (1~GeV~$<E<$~5~GeV) and high (5~GeV~$<E<$~50~GeV) energy bands.
\item The centroid of the emission moves at $\sim 1-1.5\sigma$ level toward 
that of the \veritas\ source as the energy band changes from   
$1-5$~GeV to $5-50$~GeV. The centroid is inconsistent with the PWN location, suggesting
that the PWN is not the major contributor in the present energy range. 
\item The centroid of the emission is consistent with that of \egret\ (3EG~J0617+2238), 
displaced more than $5\times \theta_{68}^{error}$(\magic) from that of \magic\ (J0610+225), 
and at $1.5\times \theta_{68}^{error}$(\veritas) that of \veritas\ (VER~J0616.9+2230). 
\item The extended source region overlaps almost completely that of \veritas. 
A group of molecular clouds (Clouds B, C, D, F, and G), the SNR shell and the PWN 
are within the overlapping region ($\theta_{68}^{ext}$), leaving possibility 
that some or all of them contribute to the observed emission.
\item The SED can not be represented by a single power-law  
but is consistent with a broken power-law with a break at $E_\gamma = 3.25 \pm 0.6$~GeV. 
\item The SED has a broad peak between a few 100~MeV and $\sim 5$~GeV 
which is consistent with the majority of the emission coming from neutral pion decays. 
For the emission being hadronic originating from a single proton population, 
the underlying proton spectrum is consistent with a broken power-law shape (chi-square 
per degree-of-freedom of 9.6/14) but not with an exponential cut-off (30.3/15).  
For the estimated total mass of interacting gas of $10^4$~M$_\odot$, 
the total energy in the pion-producing protons is estimated to be $(0.56-2.2) \times 10^{49}$~erg.
\end{itemize}

Higher statistics is needed to establish association or non-association 
of the gamma-ray emission with the molecular clouds and/or the PWN as well as 
CR injection process from the SNR into the Galactic space. 
Identification of the emission mechanisms and underlying CR spectra 
effective in individual sites will follow after such studies. 

\fermi\ LAT is expected to accumulate needed statistics well within the planned mission lifetime.   

\begin{center}
{\bf{\it{Acknowledgements}}}
\end{center}

The $Fermi$ LAT Collaboration acknowledges generous ongoing support
from a number of agencies and institutes that have supported both the development
and the operation of the LAT as well as scientific data analysis.
These include the National Aeronautics and Space Administration and
the Department of Energy in the United States, the Commissariat \`a l'Energie Atomique
and the Centre National de la Recherche Scientifique / Institut National de
Physique Nucl\'eaire et de Physique des Particules in France,
the Agenzia Spaziale Italiana and the Istituto Nazionale di Fisica Nucleare in Italy,
the Ministry of Education, Culture, Sports, Science and Technology (MEXT),
High Energy Accelerator Research Organization (KEK) and Japan Aerospace Exploration Agency
(JAXA) in Japan, and the K.~A.~Wallenberg Foundation, the Swedish Research Council
and the Swedish National Space Board in Sweden.

Additional support for science analysis during the operations phase is gratefully acknowledged from the Istituto Nazionale di Astrofisica in Italy and the Centre National d'Etudes Spatiales in France.

\newpage

\begin{deluxetable}{lc}
\tablecolumns{2}
\tablewidth{9cm}
\tablecaption{Selection for the Data Set}
\tablehead{
\colhead{Parameter}	& \colhead{Value}
}
\startdata
Time Period (MET) 	& $239557417-268416079$	\\
Energy Range ........	& $200$~MeV - $50$~GeV	\\
ROI	........................	& $\le 15^\circ$ in radius\\
Photon Class ..........	& \textit{Pass 6 Diffuse} \\
Additional Cut ........... & Zenith angle $\le 105^\circ$ \\
\enddata
\label{dataselection}
\end{deluxetable}

\begin{deluxetable}{lcccccc}
\tablecolumns{7}
\tablewidth{16cm}
\tablecaption{Centroid and Extension of the IC~443 Contribution}
\tablehead{
\colhead{Model}	& \colhead{$\ell$ ($^\circ$)}	& \colhead{$b$ ($^\circ$)}	& \colhead{$\theta_{68}^{error}$ ($^\circ$) \tablenotemark{a}}	
& \colhead{$\theta_{68}^{ext}$ ($^\circ$) \tablenotemark{a}}	& \colhead{$\theta_{95}^{ext}$ ($^\circ$) \tablenotemark{a}} & \colhead{TS$_{ext}$ \tablenotemark{b}}
}
\startdata
$1 - 5$~GeV & & & & & & \\
$\;\;\;$ Point Source & $189.05$	& $3.04$	& $0.02$	& $...$	& $...$	& $0$ \\
$\;\;\;$ Gaussian & $189.05$	& $3.05$	& $0.02$	& $0.27$~$\pm$~$0.03$	& $0.44$~$\pm$~$0.04$ & $+106/+121$ \\
$5 - 50$~GeV & & & & & & \\ 
$\;\;\;$ Point Source & $189.05$	& $2.98$	& $0.03$	& $...$	& $...$	& $0$ \\
$\;\;\;$ Gaussian & $189.06$	& $3.00$	& $0.03$	& $0.26$~$\pm$~$0.04$	& $0.42$~$\pm 0.07$ & $+81/+212$ \\
$1 - 50$~GeV & & & & & & \\ 
$\;\;\;$ Point Source & $189.05$	& $3.02$	& $0.02$	& $...$	& $...$	& $0$ \\
$\;\;\;$ Gaussian & $189.05$	& $3.03$	& $0.02$	& $0.27$~$\pm$~$0.03$	& $0.45$~$\pm$~$0.05$ & $+212/+362$ \\
\enddata
\label{extensiontable}
\tablenotetext{a}{Errors of centroids and extensions quoted include systematic errors.} 
\tablenotetext{b}{The two values shown in this column are the lower and higher of the 
TS$_{ext}$ obtained with the default PSF (the first) and worst-case PSF (the second) described in the text.} 
\end{deluxetable}

\begin{deluxetable}{lccc}
\tablecolumns{4}
\tablewidth{15cm}
\tablecaption{Spectrum of the IC~443 Contribution}
\tablehead{
\colhead{$E_{center}$[MeV]} & \colhead{$dN/dE$~[cm$^{-2}$~s$^{-1}$~MeV$^{-1}$]} & \colhead{$dN/dE$~(stat error)} & \colhead{$dN/dE$~(sys error)}
}
\startdata
247.31 & 8.63$\times 10^{-10}$ & 4.80$\times 10^{-11}$ & 6.37$\times 10^{-11}$ \\
378.19 & 4.11$\times 10^{-10}$ & 1.87$\times 10^{-11}$ & 2.52$\times 10^{-11}$ \\
578.32 & 1.72$\times 10^{-10}$ & 7.81$\times 10^{-12}$ & 8.89$\times 10^{-12}$ \\
884.36 & 8.31$\times 10^{-11}$ & 3.59$\times 10^{-12}$ & 6.11$\times 10^{-12}$ \\
1352.34 & 3.25$\times 10^{-11}$ & 1.65$\times 10^{-12}$ & 3.12$\times 10^{-12}$ \\
2067.96 & 1.51$\times 10^{-11}$ & 8.36$\times 10^{-13}$ & 1.78$\times 10^{-12}$ \\
3162.27 & 5.81$\times 10^{-12}$ & 4.06$\times 10^{-13}$ & 8.14$\times 10^{-13}$ \\
4835.67 & 2.40$\times 10^{-12}$ & 2.05$\times 10^{-13}$ & 3.90$\times 10^{-13}$ \\
7394.58 & 7.49$\times 10^{-13}$ & 8.98$\times 10^{-14}$ & 1.38$\times 10^{-13}$ \\
11307.60 & 2.80$\times 10^{-13}$ & 4.29$\times 10^{-14}$ & 5.61$\times 10^{-14}$ \\
17291.30 & 9.13$\times 10^{-14}$ & 1.93$\times 10^{-14}$ & 1.82$\times 10^{-14}$ \\
26441.41 & 2.95$\times 10^{-14}$ & 8.94$\times 10^{-15}$ & 5.90$\times 10^{-15}$ \\
40433.51 & 6.46$\times 10^{-15}$ & 3.35$\times 10^{-15}$ & 1.29$\times 10^{-15}$ \\
\enddata
\label{spectrumtable}
\end{deluxetable}

\begin{deluxetable}{lcccccc}
\tablecolumns{7}
\tablewidth{17.5cm}
\tablecaption{Spectral Fit to the IC~443 Contribution}
\tablehead{
\colhead{}	 & Model	& \colhead{$\gamma _1$}	& \colhead{$\gamma _2$}	& \colhead{E$_{break}$ (GeV)} & \colhead{F$_{200}$\tablenotemark{a} ($10^{-7}$~cm$^{-2}$~s$^{-1}$)} & \colhead{$\chi^2/dof$}
}
\startdata
IC~443	& Broken PL	& $1.93$~$\pm$~$0.03$	& $2.56$~$\pm$~$0.11$	& $3.25$~$\pm$~$0.6$	& $2.85$~$\pm 0.07$ & $8.9/9$ \\
IC~443	& PL	& $2.08$~$\pm$~$0.02$	& $...$	& $...$	& $3.00$~$\pm 0.07$ & 
$90/11$ \\      
\enddata
\tablenotetext{a}{Total flux integrated above $200$ MeV obtained with \textit{gtlike} assuming the best-fit broken power-law model 
and the best-fit 2D Gaussian spatial distribution.}
\label{spectralfit}
\end{deluxetable}

\begin{deluxetable}{lcccccc}
\tablecolumns{5}
\tablewidth{16cm}
\tablecaption{Summary of Locations and Extensions of the Gamma-Ray Sources}
\tablehead{
\colhead{Observation} & \colhead{$\ell\ (^\circ)$} & \colhead{$b\ (^\circ)$} & \colhead{$\theta_{68}^{error}$ of localization ($^\circ$)} & \colhead{$\theta_{68}^{ext}$ of extension ($^\circ$)}
}
\startdata
{\it{EGRET}} & 189.00 & 3.05 & 0.13 ($\theta_{95}^{error}$) & N/A \\
{\it{MAGIC}} & 189.03 & 2.90 & $\pm$~$0.025$ (stat) $\pm$~$0.017$ (sys) & N/A \\
{\it{VERITAS}} & 189.07 & 2.92 & $\pm$~$0.03$ (stat) $\pm$~$0.08$ (sys)
& $0.24\pm 0.05$ (stat) $\pm 0.06$ (sys)\\
{\it{Fermi}} & 189.05 & 3.03 & $\pm$~$0.01$ (stat) $\pm$~$0.02$ (sys) & $0.27$~$\pm$~$0.01$ (stat) $\pm$~$0.03$ (sys)\\
\enddata
\label{locationsummary}
\end{deluxetable}

\newpage

\setcounter{figure}{0}

\begin{figure}
\centering
\epsscale{1.1}
\includegraphics[width=12cm]{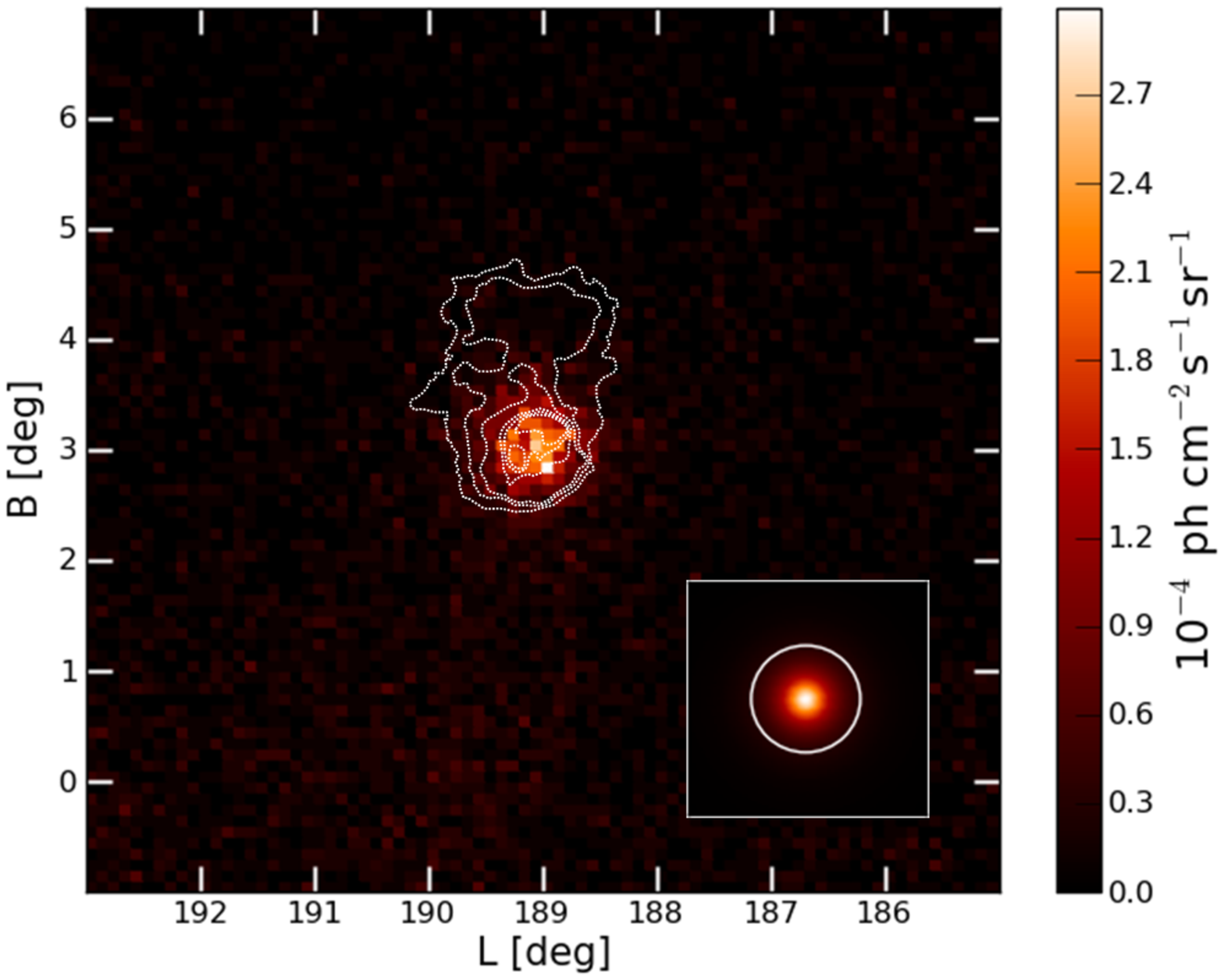}
\includegraphics[width=12cm]{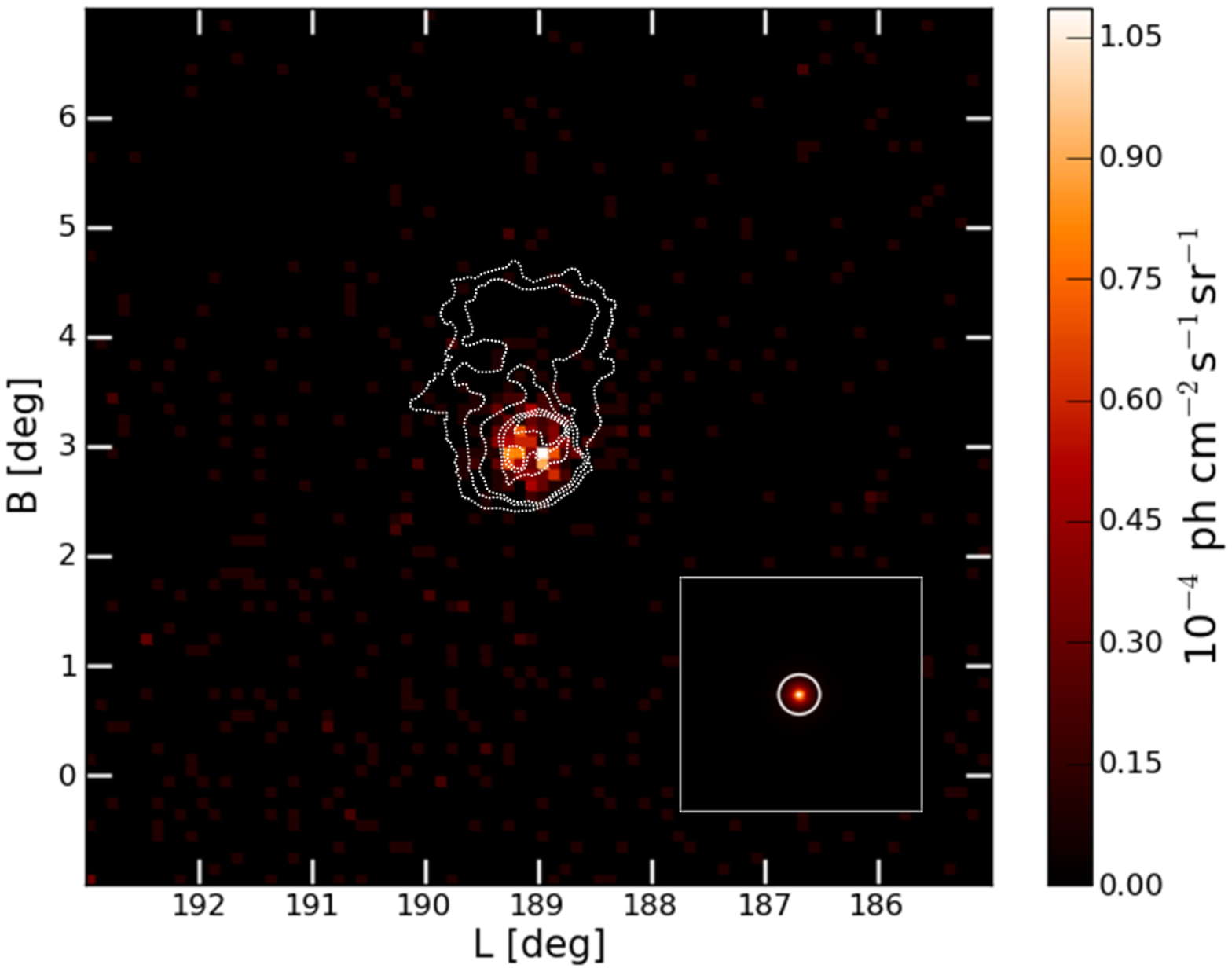}
\caption{Intensity map of the IC~443 region in the
$1-5$~GeV (left) and $5-50$~GeV (right) bands.
Units of intensity are $10^{-4}$~cm$^{-2}$~s$^{-1}$~sr$^{-1}$ for the color scale.
The overlay is the 2.7~GHz radio continuum contours taken from \citet{Furst90}.
The insets are the spectrum-weighted LAT PSF for each energy band, with the white circles
showing the corresponding $\theta_{68}^{psf}$. 
\label{intensitymapscolor}}
\end{figure}

\end{document}